\documentclass[fleqn,usenatbib]{mnras}

\usepackage{newtxtext,newtxmath}

\usepackage[T1]{fontenc}
\usepackage{ae,aecompl}

\usepackage{blindtext}

\usepackage{graphicx}	
\usepackage{multicol}
\usepackage{amsmath}	
\usepackage{amssymb}	
\usepackage{lipsum}
\usepackage{siunitx}
\usepackage[dvipsnames]{xcolor}
\usepackage{comment}
\usepackage{ulem}
\usepackage[shortlabels]{enumitem}

\newcommand{\sex}{\texttt{SExtractor}~}
\newcommand{\sexp}{\texttt{SExtractor}}
\newcommand{\btof}{\texttt{blend2flux}~}
\newcommand{\btofp}{\texttt{blend2flux}}
\newcommand{\btomtof}{\texttt{blend2mask2flux}~}
\newcommand{\unet}{U-Net~}
\newcommand{\unetp}{U-Net}

\title[Blended galaxy photometry with CNNs]{Photometry of high-redshift blended galaxies using deep learning}

\author[A. Boucaud et al.]{Alexandre Boucaud,$^{1}$\thanks{E-mail: aboucaud@apc.in2p3.fr}
Marc Huertas-Company,$^{2,3}$
Caroline Heneka,$^{4}$
\newauthor
Emille E. O. Ishida,$^{5}$
Nima Sedaghat,$^{6}$
Rafael S. de Souza,$^{7}$
Ben Moews,$^{8}$
\newauthor
Herv\'e Dole,$^{9}$ Marco Castellano,$^{10}$  Emiliano Merlin,$^{10}$ Valerio Roscani,$^{10}$
\newauthor
Andrea Tramacere,$^{11}$ Madhura Killedar$^{12}$ and Arlindo M. M. Trindade$^{13}$ \newauthor 
for the COIN Collaboration
\\
$^{1}$APC, Univ Paris Diderot, CNRS/IN2P3, CEA/lrfu, Obs de Paris, Sorbonne Paris Cit\'e, France\\
$^{2}$Instituto de Astrof\'isica de Canarias (IAC); Departamento de Astrof\'isica, Universidad de La Laguna (ULL), E-38200, La Laguna, Spain\\
$^{3}$LERMA, Observatoire de Paris, PSL Research University,CNRS, Sorbonne Universit\'es, Universit\'e Paris Diderot, F-75014 Paris, France\\
$^{4}$Scuola Normale Superiore, Piazza dei Cavalieri 7, 56126 Pisa, Italy\\
$^{5}$Universit\'e Clermont Auvergne, CNRS/IN2P3, LPC, F-63000 Clermont-Ferrand, France\\
$^{6}$Department of Computer Science, University of Freiburg, Georges-Koehler-Allee 052, 79110 Freiburg, Germany\\
$^{7}$Department of Physics \& Astronomy, University of North Carolina at Chapel Hill, NC 27599-3255, USA\\
$^{8}$Institute for Astronomy, University of Edinburgh, Royal Observatory, Edinburgh EH9 3HJ, UK\\
$^{9}$Institut d'Astrophysique Spatiale, CNRS, Univ. Paris-Sud, Universit\'e Paris-Saclay, F-91400 Orsay, France\\
$^{10}$INAF-Osservatorio Astronomico di Roma, Via Frascati 33, I-00078, Monte Porzio Catone. Italy\\
$^{11}$Department of Astronomy, University of Geneva, 24 rue du G\'en\'eral-Dufosur 1211 Gen\`eve, Switzerland\\
$^{12}$Sydney Informatics Hub, The University of Sydney, Sydney, NSW 2008, Australia\\
$^{13}$Instituto de Astrof\'sica e Ci\^encias do Espa\c{c}o, Universidade do Porto, CAUP, Rua das Estrelas, 4150-762 Porto, Portugal
}
\date{Accepted XXX. Received YYY; in original form ZZZ}

\pubyear{2019}

\begin{document}
\label{firstpage}
\pagerange{\pageref{firstpage}--\pageref{lastpage}}
\maketitle

\begin{abstract}
The new generation of deep photometric surveys requires unprecedentedly precise shape and photometry measurements of billions of galaxies to achieve their main science goals. At such depths, one major limiting factor is the blending of galaxies due to line-of-sight projection, with an expected fraction of blended galaxies of up to 50\%. Current deblending approaches are in most cases either too slow or not accurate enough to reach the level of requirements. This work explores the use of deep neural networks to estimate the photometry of blended pairs of galaxies in monochrome space images, similar to the ones that will be delivered by the Euclid space telescope. Using a clean sample of isolated galaxies from the CANDELS survey, we artificially blend them and train two different network models to recover the photometry of the two galaxies. We show that our approach can recover the original photometry of the galaxies before being blended with $\sim7\%$ accuracy without any human intervention and without any assumption on the galaxy shape. 
This represents an improvement of at least a factor of 4 compared to the classical \sex approach. We also show that forcing the network to simultaneously estimate a binary segmentation map results in a slightly improved photometry. All data products and codes will be made public to ease the comparison with other approaches on a common data set. 
\end{abstract}

\begin{keywords}
galaxies: general -- galaxies: photometry -- methods: data analysis -- methods: statistical --  techniques: image processing
\end{keywords}



\section{Introduction}
\label{sec:intro}

The upcoming years will be marked by the arrival of a new generation of deep and wide galaxy surveys from ground, \citep[e.g., Large Synoptic Survey Telescope (LSST),][]{Ivezic2008}, and space \citep[e.g, Euclid,][]{Euclid2016}. Under this new paradigm of \textit{big-data} surveys, the community aims to achieve an unprecedented level of accuracy and precision both in terms of photometry \citep[e.g., photometric redshifts,][]{Krone-Martins2014,Elliott2015,Beck2017,Salvato2018}, and  shear measurements \citep[e.g.,][]{Kilbinger2017,Kitching2017} for an unprecedented number of objects. This requires to revisit most of the commonly used procedures to extract measurements from images, in order to reduce as far as possible all the systematic effects and reach the requirements. One particular important source of error is the blending of sources. As surveys become deeper and deeper, we expect an increasing fraction of overlapping galaxies which could bias the measurements at levels beyond requirements \citep{Dawson2015}. For example, the estimates for LSST say that $\sim$45\%-66\% of the sources are expected to overlap to a degree of being problematic for a number of methods, with $\sim$75\% of blends probably composed of only two objects \citep{dawson_2014, Dawson2016}. If galaxies are not properly separated, their photometry is biased, which has a direct impact on the derived redshift (and all other physical properties). Efficient algorithms to automatically separate (deblend) detected sources are crucial and will be a key ingredient in the processing pipelines of the next generation surveys. However, there is currently no standard solution in the literature and deblending remains an open issue among the community. 

The most widely used software for detecting and separating objects in large fields is \texttt{SExtractor}\footnote{\href{https://www.astromatic.net/software/SExtractor}{https://www.astromatic.net/software/SExtractor}} \citep{SExtractor} but its use is far from being an optimal solution. In a nutshell, this software looks for wells in the luminosity profiles of galaxies using multiple thresholds. The main problem with such an approach is that it is very sensitive to the configuration parameters and it is difficult to configure so that it works in a wide variety of cases. The fraction of blended sources which are not identified as such can reach significant fractions \citep{Laidler2006}. An alternative way is to simultaneously fit a parametric model to all galaxies in the image and use the best fit models to estimate the photometry \citep{Pignatelli2006, Mancone2013}. This approach typically reaches better photometric accuracy but still requires to properly identify the centroids of all the different objects. It also assumes simplistic models for the galaxy surface brightness distribution which do not encapsulate all the diversity of galaxy morphologies, especially in the more distant Universe. It is also expensive in terms of computing time. The classical deblending approaches are therefore insufficient to reach the level of requirements on measurements of galaxy properties for upcoming surveys.  It is thus timely to investigate and compare different approaches.

 Several groups are working on alternative solutions more adapted to large volumes of data \citep{Joseph2016, Tramacere2016, Ivezic2017}. For example, recent works by the LSST collaboration \citep{Melchior2018} have started to develop more global approaches based on non-negative matrix factorisation, that can achieve a more efficient source separation and enable to put flexible constraints or priors on the shape of the signals. This approach is however optimised for ground based data in which galaxies have little resolved structures and also takes full advantage of the multi-wavelength nature of LSST data. It is less well adapted for monochrome space data such as the images delivered by the Euclid space telescope \citep{Jones2019}. 

The goal of this paper is to explore if machine learning and more precisely deep learning is an approach worth investigating for segmenting blended galaxies and estimating their photometry.  During recent years, the use of deep learning approaches for tasks related to galaxy images has become a burgeoning field of research in astronomy. One of the earliest and most pervasive area of application is the classification of galaxy morphologies \citep[e.g.,][]{Dieleman2015, Barchi2017, DominguezSanchez2018, HuertasCompany2018, Khalifa2018}. 
More recent research includes the recovery of galaxy features in noisy images by \citet{Schawinski2017}, the finding of galaxy-galaxy strong lensing effects by \citet{Lanusse2018}, 
and the generation of physically realistic synthetic galaxy images to augment existing data sets and consequently provide the aforementioned deep learning approaches with larger training sets \citep{Ravanbakhsh2017, Fussell2019}.

In a recent work, \citet{Reiman2019} used deep learning for the first time to deblend SDSS galaxies. They introduce a modified Generative Adversarial Network \citep[GAN,][]{Goodfellow2014} to separate blended galaxies, combining aspects of the super-resolution GAN (SRGAN) by \citet{Ledig2017} and the deep residual learning framework by \citet{He2016}. With the generator as a modified residual network that features two branches, each branch generates one of the two blended galaxies. They show promising results. However, their procedure to generate blended images for training based on the pixel-wise maximum of the two individual stamps does not reflect the true process resulting in line-of-sight blending which sums the photons coming from both sources. 

In this paper we further explore the use of machine learning to both segment and measure the photometry of blended pairs of galaxies. The approach presented here is designed having Euclid data in mind as the main target of application (i.e. monochrome space based data). The goal is thus to obtain a neural network optimised to predict the photometry of pairs of galaxies observed with fairly high spatial resolution in one single band.

This paper is organised as follows: 
In Section~\ref{sec:data},
we describe the realistic image data set of blended galaxy pairs we created. We then detail the reasons behind the choice of deep learning methods for this paper and carefully unroll the methodology used to set up our networks in Section~\ref{sec:strategy}. These methods are applied to our emulated data set in
Section~\ref{sec:results}, where we compare the results with \sexp, before discussing the pros and cons in the final Section~\ref{sec:conclusion}.


\section{Data set of artificially blended galaxies}
\label{sec:data}

\begin{figure*}
	\centering
	\includegraphics[width=0.9\textwidth]{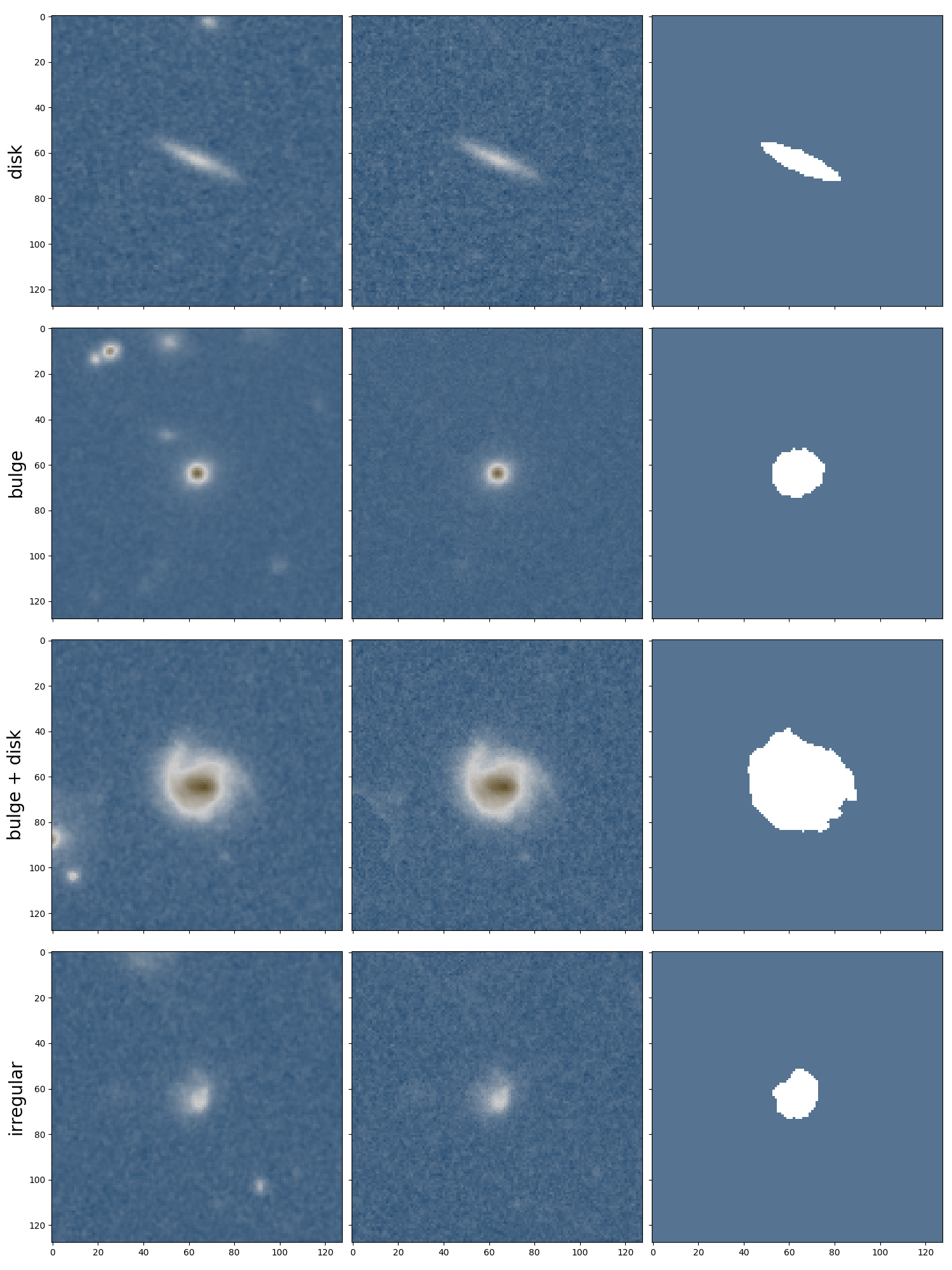}
    \caption{Selection of CANDELS cut-outs displaying at their centre galaxies with different morphologies. The left column shows the original CANDELS image. The middle one shows the same cutout after the neighbours removal procedure, leaving the central galaxy fully isolated. The rightmost column shows the \sex segmentation map for that isolated galaxy. The galaxy images have been asinh-stretched to enhance the details.}
  \label{fig:blending}
\end{figure*}

Galaxy blending is a confusion effect created by the projection of photons from galaxies on a given line-of-sight, to the 2D plane. As telescopes get more sensitive, we have access to a higher number of galaxies and thus to a higher chance of finding multiple objects in the same line-of-sight \citep{dawson_2014}.

The quantification of the effects of blending on the derived galaxy properties is a difficult task by nature, due to the integration of photons by our sensors and the intrinsic convolution by the point spread function of the instrument. Most existing methods require additional knowledge (several wavelength bands), or a priori knowledge, like parametric models, of the galaxy profiles, symmetries, etc.  Moreover, to assess the accuracy of such methods, we are often left with bottom-up approaches like the simulation of galaxy blending using software like \texttt{GalSim} \citep{galsim}, for which we have access to the true light distribution of each object in the image. But as realistic as they can be, simulated images often show their limits when compared with the diversity and the singularity of real data images \citep{Haussler2007}. This is particularly critical for machine learning which implicitly assumes that the training sets are fully representative of the real data.

In order to get a realistic representation of observations, for this work we decide to simulate blended objects from real observations. Although this approach eventually propagates the biases and errors existing in the observations, it has the advantage of including fully realistic morphologies. We describe in the next paragraph the methodology we follow to generate our galaxy sample.

\subsection{Parent Sample}

The parent sample used in this paper is the H-band selected catalogue from the Cosmic Assembly Near-infrared Deep Extragalactic Legacy (CANDELS) survey, presented in \citet{DiMauro2018}. The catalogue contains galaxies with $F_{160W}<23.5$, for which both visual morphologies and parametric bulge-disc decomposition are performed. From this parent data set, we first define a clean sub-sample of isolated galaxies with unambiguous morphologies that are then used to perform the blends. More precisely, we use the neural-network-based morphological classification published in \citet{HuertasCompany2015} and select galaxies with four different morphological types: 
\begin{itemize}
    \item pure bulges:  $P_{\rm SPH}>0.8$\,,
    \item pure disks: $P_{\rm DISK}>0.8$\,,
    \item two component bulge + disk: $P_{\rm SPH}>0.8$ \& $P_{\rm DISK}>0.8$\,,
    \item irregular galaxies: $P_{\rm IRR}>0.8$\,.
\end{itemize}

Note that the purpose of this selection is not to have a complete sample of galaxies, but to have a clean data set of isolated galaxies with different morphologies for which we can reasonably trust the segmentation procedure. By selecting galaxies with very large probabilities of being in a given morphological type we can be reasonably certain that we remove originally blended systems or complex structures such as mergers. 

From this initial sample, we generate $128\times128$ pixel stamps centred on the objects. We then remove all other objects present in the stamps. To that purpose, we apply a morphological dilation to the original segmentation obtained with \sex and replace all distinct regions with random pixels sampled from empty regions in the background. This process is illustrated in Figure~\ref{fig:blending}.

In order to further clean the sample, we visually inspect the selected galaxies and remove the ones which still present anomalies such as originally blended systems not detected by \sexp, or the ones for which the removal of companions created some visual artefacts in the images. The final sample results in $\sim2,000$ galaxies whose types are summarised in Table~\ref{tab:galprops}. Figure~\ref{fig:blending} shows a selection of these galaxy stamps, along with the stamp after the removal of neighbouring objects and the associated \sex segmentation map of the central isolated galaxy.

\begin{table}
	\centering
	\caption{Morphological mix of the final dataset used to generate the blended systems.}
	\label{tab:galprops}
	\begin{tabular}{lcc}
		\hline
		Galaxy type & \# before inspection & \# after inspection \\
		\hline
		bulge           & 386 & 352 \\
		disk            & 473 & 433 \\
		bulge + disk    & 884 & 702 \\
        irregular       & 875 & 432 \\
		\hline
        TOTAL           & 2618 & 1919
	\end{tabular}
\end{table}

\subsection{Blending}

\begin{figure*}
	\centering
	\includegraphics[width=0.9\textwidth]{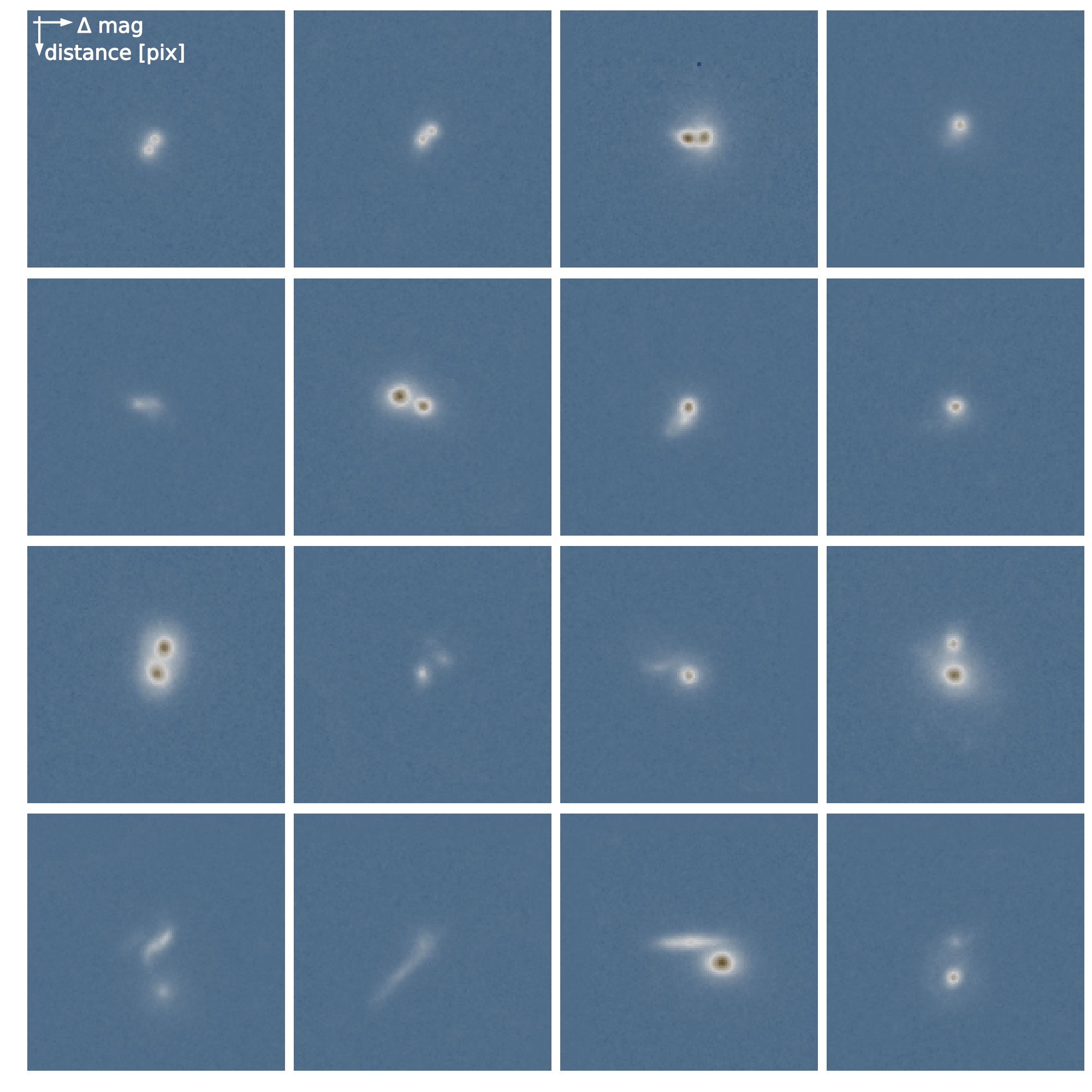}
    \caption{Selection of blended systems created and used in this work. The stamps are ordered vertically by the distance in pixels between the galaxy centres, and horizontally with respect to the magnitude difference between the galaxies. The images have all been asinh-stretched for visualisation purposes.}
  \label{fig:blends}
\end{figure*}

To create the artificially blended systems, we combine the galaxies of the clean sample we just obtained using the following procedure.
First, we randomly select one galaxy, referred to as the \textit{central galaxy}, with a magnitude and an effective radius  respectively denoted $\rm mag_{cen}$ and $R_{\rm cen}$. $R_{\rm cen}$ is the semi-major axis of the best Sersic fit model from the catalog by~\cite{DiMauro2018}. Second, we pick a second galaxy in the catalogue, referred to as the \textit{companion galaxy} with properties $\rm mag_{comp}$ and $R_{\rm comp}$, so that it satisfies $\rm mag_{cen} - 2 < mag_{comp} < mag_{cen} + 2$. 
Then we set $R = \text{max}(R_{\rm cen}, R_{\rm comp})$ as the biggest effective radius between the two galaxies and randomly select a couple of shifts ($\Delta x$, $\Delta y$) from a uniform distribution ranging from $0.5\cdot R$ and half of the image size.
We use these shifts to apply a translation to the stamp of the \textit{companion galaxy}.
Finally, the blend is created by adding up the pixels of two stamps.

Note that the blending process contains two over-simplifications as compared to real observed blends. Firstly, we avoid overlap in the very inner parts of the central galaxy ($<0.5R_e$) and secondly, the central galaxy is always placed at the centre of the stamp. We are fully aware of these simplifications but consider this enough complexity for our blends in a first proof-of-concept work. 

We repeat this process to build up a sample of $30,000$ blended galaxies, which necessarily contains some redundancy because each galaxy appears in multiple stamps. However since there are enough degrees of freedom coming from the selection of the companion and the shifts, this redundancy is not to be considered problematic. It allows us to build a large enough sample to train the networks as described in the following. We show in Figure~\ref{fig:blends} some examples of blended pairs with different magnitude differences and distances between the two galaxies.

To summarise, at the end of this process, we have for every generated blend system:
\begin{itemize}
  \item[-] the original CANDELS cut-outs of the \textit{central} and \textit{companion galaxy},
  \item[-] the associated \sex segmentation maps,
  \item[-] the associated \sex photometry (\texttt{FLUX\_AUTO}),
  \item[-] the generated blended  stamp.
\end{itemize}

With the purpose of triggering the comparison with other approaches, the software used to generate the blends as described above has been publicly released as a package called \texttt{candels-blender}.\footnote{\url{https://github.com/aboucaud/candels-blender}}


\subsection{Training, Validation and Test data sets}
As explained in the previous sections, the blend stamps contain some level of redundancy since the same galaxy can appear in several of them. This could artificially improve the results evaluated in the test set because the network might have seen already the same galaxy in the training phase. To avoid this potential bias, we adopt a specific procedure. Following a standard approach in machine learning \citep[e.g.][]{Bengio2012}, we split the dataset into three subcategories: training, validation and test, respectively 60\%, 20\% and 20\% of the full dataset. During the training, the model loss (i.e. cost function) is periodically computed on the validation sample to ensure it is not diverging from the training, which would indicate over-fitting or a bad convergence of the network. Training and validation samples can be randomly selected from the same dataset, however the test sample, on which the metrics are computed, must be carefully chosen to be both distinct from and representative of the sample used for the training and validation. To achieve this feature and obtain meaningful results, we isolate the sample of galaxy stamps used for the test dataset at the very beginning by randomly picking them out of the catalogue. This way, all the galaxies used to construct the blends for the training and validation are never to be found in the test sample of blends, and vice-versa. In the end, we have a training/validation set composed of $25,000$ blends and a test set of $5,000$ blends. 
This generated data set is used to train several deep neural network architectures as described in the next section.

\section{Methods}
\label{sec:strategy}

\begin{figure*}
\centering
\includegraphics[width=0.95\textwidth]{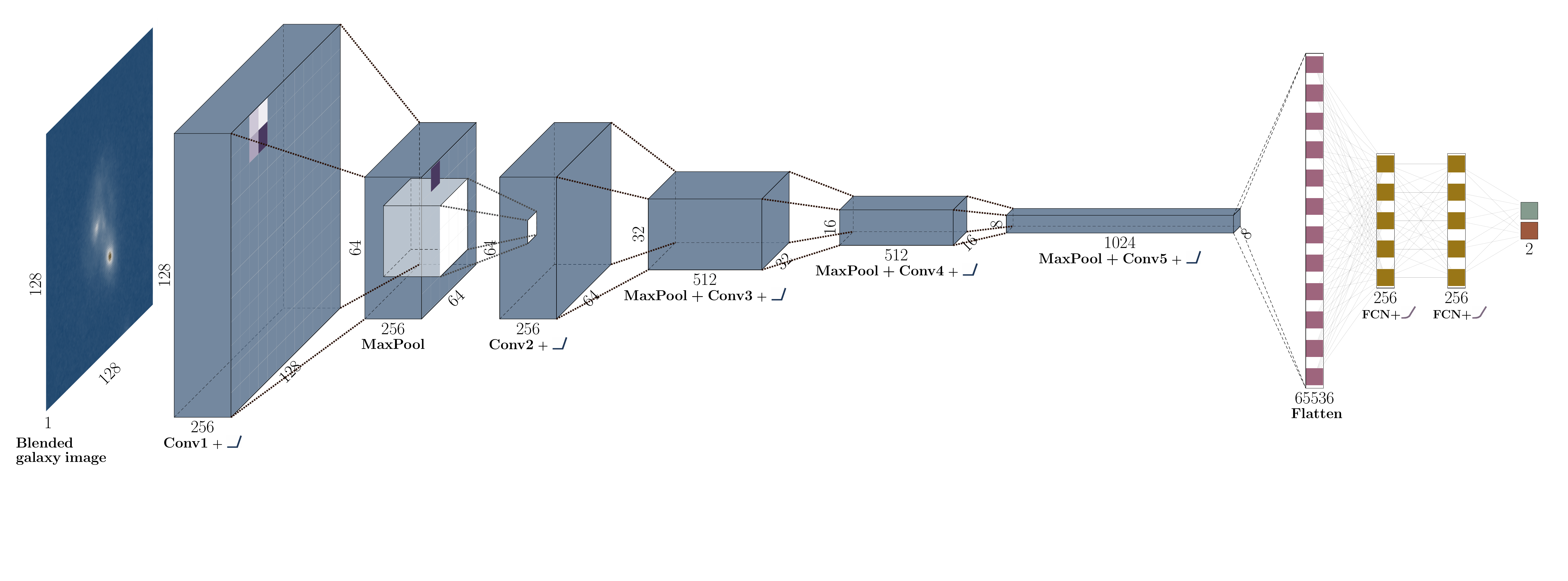}
\caption{Schematic representation of the fiducial \btof network. The network takes as input an image of a blended system and outputs the fluxes of the two galaxies. The blue boxes correspond to the convolutional part of the network. The yellow part is the fully connected section. The sizes of the different layers and convolutions are also indicated.}
\label{fig:seqstack}
\end{figure*}

\begin{figure*}
\centering
\includegraphics[width=0.85\textwidth]{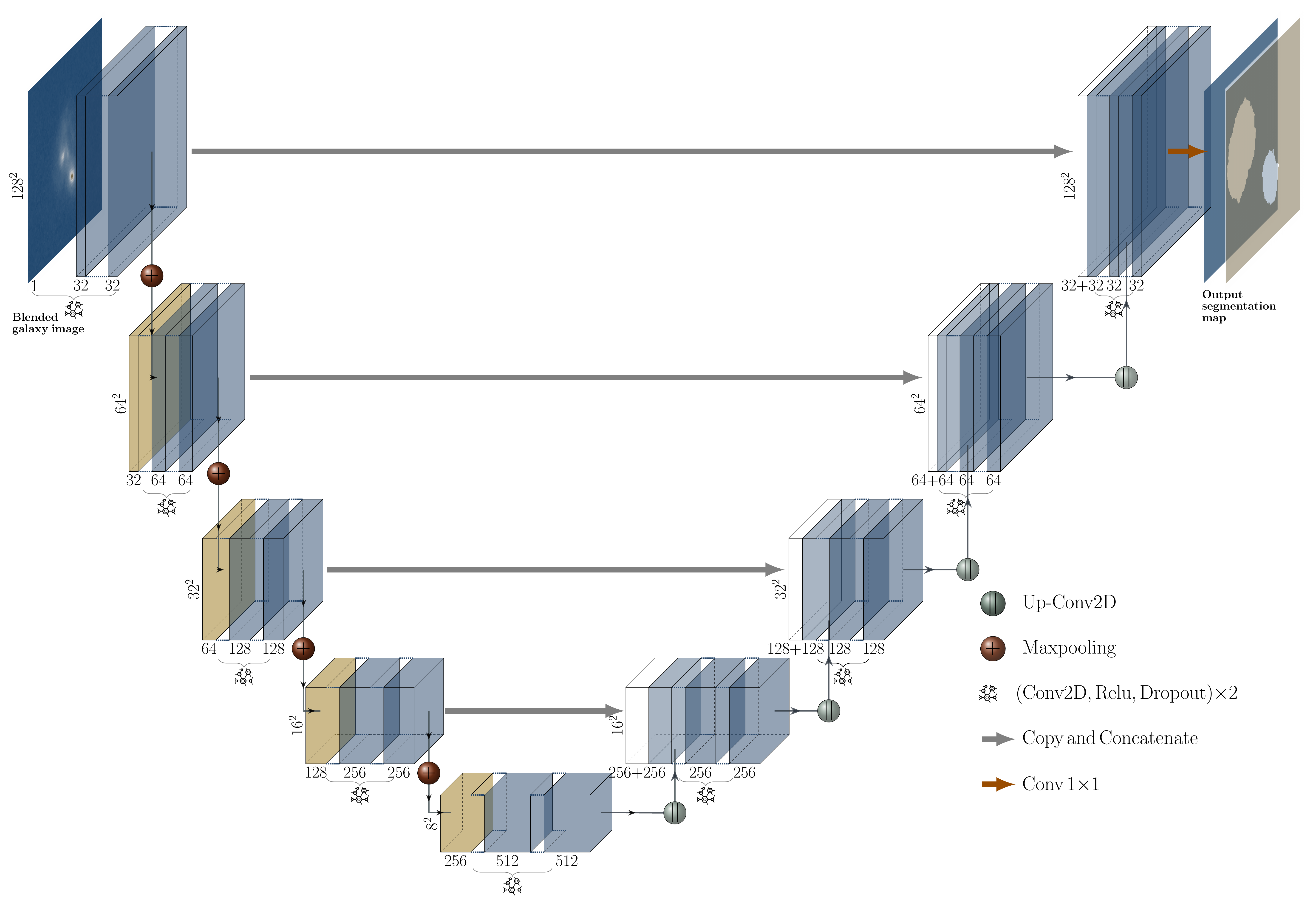}
\caption{Schematic representation of the \unet part of the \btomtof network. The network takes as input an image of blended system and outputs a segmentation map. The lines indicate the connections among the different layers.}
\label{fig:unet}
\end{figure*}

Our goal is to recover, with deep learning, the photometry of the two galaxies before the blending process. The sample being made of real galaxies, we make the assumption that the ground truth (also referred to as \textit{the target} in supervised learning) is the flux of the isolated galaxy computed by \sex on the original CANDELS cut-out. We also assume that the segmentation mask provided by \sex for the isolated galaxies is correct. We understand that these are strong assumptions. 
However, the main purpose of the work does not depend on the absolute accuracy of the training sample. The main objective is to calibrate how well we can recover the photometry of blended galaxies relative to the accuracy obtained on the same galaxies when they are isolated. In that respect, the ground truth can be replaced with any other measurement.

We perform two different experiments. In the first one we use a standard Convolutional Neural Network \citep[CNN,][]{LeCun1989, Schmidhuber2015, Sze2017} to directly compute the fluxes of the two galaxies from the blend image. We call this configuration \btof. In the second experiment, we recover with a unique architecture, both the segmentation maps and the fluxes of the two galaxies. The idea is to calibrate whether having information on the segmentation map helps the network to  obtain a more reliable photometry. We call this second experiment \btomtof.

The networks are implemented, trained and evaluated using the Python API \texttt{Keras},\footnote{https://keras.io} which runs on top of \texttt{TensorFlow}.\footnote{https://github.com/tensorflow/tensorflow} The source code needed to reproduce the results of this paper as well as all the plots will be publicly released upon acceptance.

\subsection{Configuration 1: blend2flux}
\label{sub:b2f}

Experience with deep learning has proven that reducing pre-processing to a minimum often results in better results \citep{Liang2015}.  We thus start off with a deep neural network model that predicts fluxes directly from the blended images without any intermediate step. We use to that purpose a standard CNN configuration including a feature extraction convolutional part followed by a fully connected (or dense) network. The input of the network is thus a 1-channel image with two blended galaxies and the output is a vector of two floating numbers corresponding to the fluxes of each galaxy. 

We build a modular version of this sequential network, where the number of layers of both the convolutional and the dense network, as well as their filter size are adjustable. The architecture whose results are shown in this paper is sketched in detail in Figure~\ref{fig:seqstack}. The CNN part is made of five convolutional layers activated using a ReLU function and using convolution kernels of size $3\times3$ only. Max-pooling layers are inserted in between each convolution layer to downsample the images. The first layer starts with a filter size of 256, and doubles this filter size every other layer. After the fifth convolutional layer, the data is flattened to be fed to a three-layer classical neural network, finally yielding a vector of size two with the fluxes.
Given that our network is aiming at correct relative flux measurements, we choose to use the mean absolute percentage error (MAPE, see equation \ref{eq:mape}) as our loss function. To adjust the weights during training, we select the Adam algorithm, a popular optimiser for deep learning due to its fast and effective learning \citep{2014arXiv1412.6980K}. Adam is an extended stochastic gradient descent algorithm, meaning it iteratively updates network weights with individual adaptive learning rates based on both first and second moments of the gradients.

\begin{equation}
    \text{MAPE}\left(y_{meas}, y_{true}\right) = \frac{100}{n} \sum_n \left|\frac{y_{true} - y_{meas}}{y_{true}}\right|
    \label{eq:mape}
\end{equation}

This \btof network, which has about $25.7$ million free parameters, is then trained from scratch using the training set of $25,000$ images. We consider the network as having converged after the validation loss, computed on the validation part of the training sample, stays on a plateau for a full ten consecutive epochs after having decreasing the learning rate several times \citep{Yao2007}. For this network, it happened after 70 epochs which took less than five hours of training on an Nvidia K80 GPU.

\begin{table}
    \centering
    \caption{\btof network performance computed on the entire test set using mean absolute percentage error (MAPE).}
    \begin{tabular}{lp{1.0cm}p{1.0cm}p{1.0cm}} 
    \hline
      Initial Filter Size & 64 & 256 & 512 \\
     \hline
     No. parameters [Mio.] & 1.6 & 25.7 & 102.7 \\
     Flux error central [\%] & 9.33 & 8.39 & 8.25 \\
     Flux error companion [\%] & 8.79 & 8.01 & 7.98 \\
     Total flux error [\%] & 9.06 & 8.20 & 8.12 \\
     \hline
    \end{tabular}
    \label{tab:btofnets}
\end{table}

The network built being modular, we trained a few variations around the fiducial network presented above to compare their relative performance. The results of the various network models as a function of the number of filters for the first convolutional layer are summarised in Table~\ref{tab:btofnets} with the fiducial results in the middle column. The table shows that doubling the initial filter size (right column) only slightly increases the performance on the validation set  regarding the estimated fluxes in Section~\ref{sec:results}, at the expense of quadrupling the number of parameters (hence the training time and computation cost). Using instead a smaller network with an initial filter size of 64 (left column) reduces the number of parameters to about 1.6 million, which has a higher impact on the performance ($\sim1\%$ worse). The network still reaches though a precision below 10 percent on estimated fluxes, despite being significantly reduced in size. We therefore want to stress here that smaller and simpler networks than our fiducial one still outperform traditional methods.

\subsection{Configuration 2: blend2mask2flux}
\label{sub:b2m}

In a second experiment, we aim at recovering the individual segmentation maps for the two galaxies in addition to the photometry. The objective of this exercise is to quantify if the segmentation maps contain additional information that the networks can use to improve the photometry. 
We achieve this objective using a concatenation of two different networks, one to produce the segmentation maps, and a second to predict the fluxes from the segmentation maps and the blend image. We call that composite network \btomtof. One important constrain when building this network was to ensure it had approximately the same number of free parameters as the fiducial \btof. 

To produce the segmentation maps, we use a suitable deep network architecture from the literature called a \unet \citep{Ronneberger:2015aa}. \unet was designed to perform bio-medical image segmentation and has already proven useful to detect and segment overlapping chromosomes. The network architecture is quite unique and characterised by an ability to capture both fine and large scale information of the input image by keeping a copy of each downsampling step (convolution + \textit{max-pooling}) and concatenating it at the upsampling step. 
For our purpose, we create a modular version of the original \unet architecture made of blocks of two  convolutional layers activated with ReLU, followed by either a downsizing or upsizing layer (respectively, max-pooling and up convolution layers). Because the output images are of the same shape as the input blend, each downsizing block is associated with an upsizing one in the network, and the model can therefore be parametrised by the number of consecutive downsizing blocks, as well as the size of the filters (number of convolution kernels). After some tests and with a range of these parameters, we selected a \unet with a depth of 5 and an initial filter size of 32, which we also refer to as the \textit{fiducial} model. The exact architecture of this segmentation network is depicted on Figure~\ref{fig:unet}. The last activation of the model is a sigmoid function that creates output images with pixels in the range $[0, 1]$. These pixels are then thresholded to obtain segmentation maps with binary values ${0, 1}$ and we use a binary cross-entropy loss to train the model. Further results of this pure segmentation stage will be discussed in a specific section \ref{sub:segmaps} at the end of this paper.

The second part of this composite model is the retrieval of the photometry using the blend image and the segmentation maps obtained with the \unetp. For this part, we use an architecture similar to the \btof model shown in Section~\ref{sub:b2f} with a reduced number of free parameters, and changing the input to a 3-channel input 
- the concatenation of the blend image, the segmentation of the central galaxy and the segmentation of the companion galaxy - (instead of 1-channel - the blend image - in the original \btof network). 
Like the \btof model, the output of the network is evaluated using the mean absolute percentage error loss. 

The composite \btomtof network is trained following a particular process. First, the \unet is trained alone to produce accurate segmentation maps of the two galaxies. Then we load the pre-trained weights of the \unet into the \btomtof network, 
and train the network end-to-end with respect to the flux retrieval, i.e. using the mean absolute percentage error loss on the photometry. Note that we still keep the loss on the segmentation part but with a weight of 0.1 compared to the photometry loss. This last optimisation step, during which we optimise the network with respect to both the segmentation and the photometry loss, also fine-tunes the segmentation maps for flux measurement. A more detailed discussion on this aspect can be found in Section~\ref{sub:segmaps}. 

This \btomtof network presented above has $18.5$ million free parameters, a number very close and even inferior to the \textit{fiducial} \btof model. The \unet part is trained from scratch on the $25,000$ image training set for about 50 epochs. Then the end-to-end \btomtof network is trained during a few hundred epochs with a small learning rate. This full process takes about fifty hours of training on a Nvidia K80 GPU, much longer than that of the \btof network. Both the model complexity, and the training process (reduced batch size for the \unet training) are accountable for that order of magnitude time difference.

\subsection{Baseline: SExtractor}
\label{sub:sex}

In order to have a baseline to compare with, we also run a classical \sex segmentation procedure on the blended systems. We highlight that the comparison is not completely fair since \sex does not only measure photometry but also detects the objects without any prior on the number of existing objects. However, the two deep learning approaches implicitly incorporate a prior on the number of blended galaxies through the training set (networks are trained only with images containing two objects).  

In order to minimise that effect, and help \sex as much as possible, we adapted the procedure reported by \citet{Galametz2013}, where \sex is first ran in a \textit{cold} mode, aiming to select the larger elements in a blended image followed by a second round where it is ran in a \textit{hot} mode - which is more sensitive to small structures. In our particular case, where the data is known to have only two elements, the \textit{cold} mode was used to scan all the images and a subsequent run with the \textit{hot} mode was restricted to those images for which \sex identified only 1 object. Our code used the Python package \texttt{sep} \citep{sep} and the parameters used for both modes are described in Table~\ref{tab:SExtractor}. 

Following this procedure, results can be divided in 3 cases:
\begin{enumerate}
    \item \sex detects exactly two galaxies (75\%): fluxes were associated with central or companion galaxy based on the closest detection.
    \item \sex detects only a single object, meaning it is not able to deblend the pair (22\%):  detected object was associated with the central galaxy if its measured centroid is located within $0.5 R_{\rm cen}$ from the centre of the image. Otherwise, detection was associated with the companion galaxy.
    \item \sex over-deblends and detects more than two objects (3\%): the two brightest detections were considered - others were ignored.  
\end{enumerate}

\begin{table}
    \centering
    \caption{\sex parameters for \textit{hot} and \textit{cold} modes.}
    \begin{tabular}{lcc} 
    \hline
     Parameter & Hot & Cold \\
     \hline
     Detection threshold & 4 & 5 \\
     Minimum pixel area per object & 6 & 10 \\
     Minimum contrast ratio &  0.0001 &  0.01 \\
     Number of thresholds for deblending & 64 & 64 \\
     \hline
    \end{tabular}
    \label{tab:SExtractor}
\end{table}

\section{Results}
\label{sec:results}

\begin{figure*}
\centering
\includegraphics[width=0.98\textwidth]{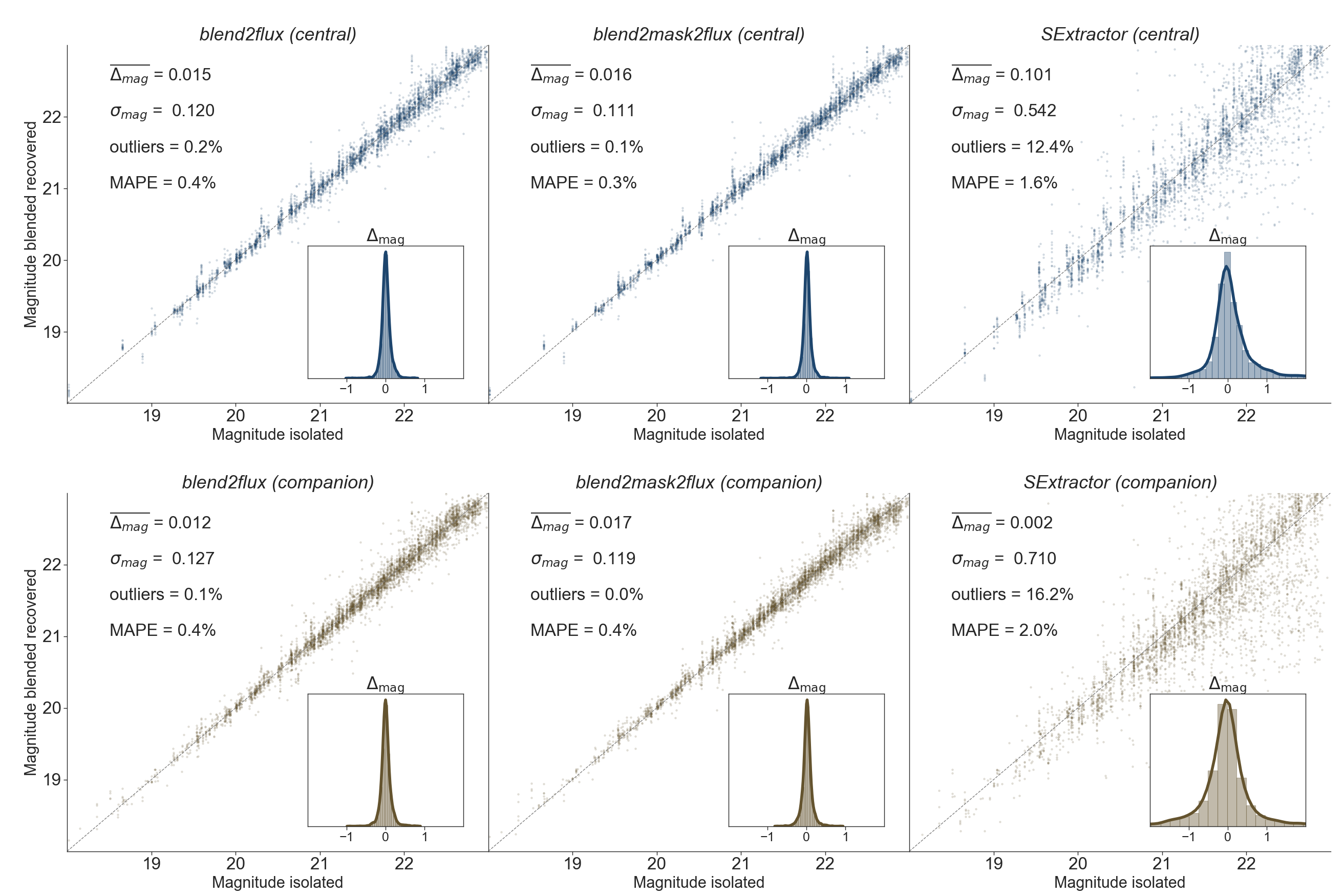}
\caption{Magnitude measured on the blend systems as a function of the magnitude measured by \sex on the same isolated galaxies (isolated magnitude). The top row shows the results for the central galaxy using the blends for which \sex detected either the two galaxies or only the central one. The bottom row shows the results for the companion galaxy using the blends for which \sex detected either the two galaxies or only the companion. The columns refer to different codes or models applied to the blend images, respectively from left to right \btofp, \btomtof and \sexp. 
The dashed line denotes identical estimation from blended and isolated galaxy images to guide the eye. 
The inner panels show the histograms of photometric errors ($\Delta_{\rm mag}={\rm mag_{blend}-mag_{isolated}}$). The numbers in each panel indicate the average photometric error $\overline{\Delta_{\rm mag}}$, the dispersion $\sigma_{\rm mag}$, the fraction of outliers, defined as $|\Delta_{\rm mag}| > 0.75$, and the mean absolute percentage error (MAPE) on the magnitude.}
\label{fig:one2one_plots}
\end{figure*}

In this section, we evaluate the results of the three experiments described previously. The main objective is to test the photometric accuracy of blended objects as compared with the photometry obtained on the same objects when they are isolated. We use the magnitude difference as the main indicator of accuracy and explore the results as a function of two main parameters: the magnitude difference between the two galaxies and the distance between the two galaxy centroids.

\subsection{Overall photometric accuracy}
\label{sub:photoacc}

\begin{figure*}
\centering
\includegraphics[width=0.98\textwidth]{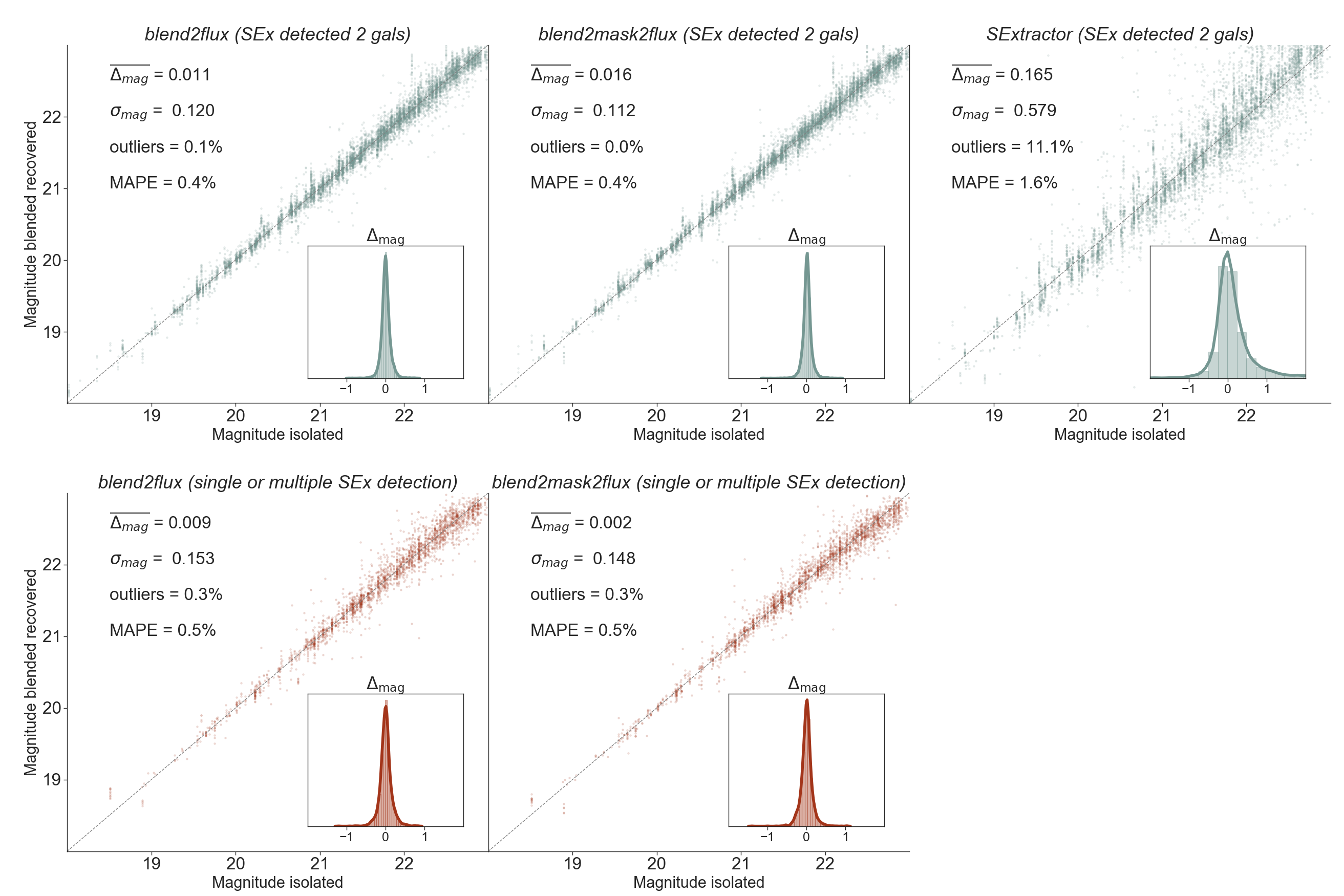}
\caption{Magnitude measured on the blend systems as a function of the magnitude measured by \sex on the same isolated galaxies (isolated magnitude). The top row shows the results for the central and companion galaxies on the blends for which \sex detected exactly two galaxies while the bottom row show the results on the blends for which \sex detected either one or more than two galaxies (under- or over-deblending). The different columns indicate different codes or models applied to the blend images, from left to right \btofp, \btomtof and \sexp. 
The dashed line denotes identical estimation from blended and isolated galaxy images to guide the eye. 
The inner panels show the histograms of photometric errors ($\Delta_{\rm mag}={\rm mag_{blend}-mag_{isolated}}$). The numbers in each panel indicate the average photometric error $\overline{\Delta_{\rm mag}}$, the dispersion $\sigma_{\rm mag}$, the fraction of outliers, defined as $|\Delta_{\rm mag}| > 0.75$, and the mean absolute percentage error (MAPE) on the magnitude.}
\label{fig:one2one_nondet}
\end{figure*}

Figures~\ref{fig:one2one_plots} and \ref{fig:one2one_nondet} show the recovered magnitude in the blended systems (hereafter output magnitude) for the three different methods, the \btof and \btomtof networks and \sexp, as a function of the magnitude measured on the same isolated galaxies (hereafter input magnitude). On Figure~\ref{fig:one2one_plots} we focus the results on the central (top) and the companion (bottom) galaxy using the blends for which \sex detects them. On Figure~\ref{fig:one2one_nondet}, we aggregate the results on both galaxies, and distinguish the cases for which \sex detects the pair (top) and over- or under-deblends (bottom). 

On both figures, the deep learning architectures behave very similarly. The relation between the two quantities is centred on the one-to-one line and the typical scatter is $\sim0.1$ magnitudes. The scatter is roughly constant over all the luminosity range explored which means that the photometry can be recovered with similar accuracy for bright and faint objects in our sample. This is clearly not the case for the \sex results which present a noticeable increase of the scatter at the faint end. This difference highlights an important advantage of a machine learning approach. If the training set is representative of the real data, the algorithm optimises the loss for all objects equally. 

In each panel of Figures~\ref{fig:one2one_plots} and \ref{fig:one2one_nondet}, we quantify in more detail the bias and scatter on the recovered photometry. The embedded histograms show the distribution of photometric error $\Delta_{\rm mag}={\rm mag_{blend}-mag_{isolated}}$ between the output and input magnitudes. The distributions for both the central and the companion galaxy are generally well centred around zero for the three codes, which indicates that the  estimators are globally unbiased. We note that the \sex panels present a slightly skewed histogram and positive bias of $0.1$ mag for the central galaxy. We explain this slight bias by looking at the selection process of the companion galaxy described in Section~\ref{sec:data}, which is skewed a bit towards selecting fainter galaxies than the central ones.

The visible difference between the methods are shown in the scatter. Both deep learning approaches present a very low scatter of $\sim0.1$ mag compared to the $\sim0.5-0.7$ mag scatter of \sexp. Another good indicator of the model performance, used for training the models, is the mean absolute percentage error (see equation \ref{eq:mape}), computed here on the magnitude. Again, both network model show good and similar performance, with always a slight advantage for the \btomtof, whereas \sex is distanced. These two indicators show an  overall improvement of the measured photometry of a factor 4 using the deep learning models compared to using \sex on these blended galaxies.

Another important difference between the methods is the fraction of catastrophic errors, i.e. cases in which the estimated photometry in the blended systems significantly differs from the input value. We arbitrarily set the threshold value to define catastrophic errors to $|\Delta_{\rm mag}| > 0.75$, which corresponds to an error of a factor of 2 in flux. The fraction of outliers defined that way is two orders of magnitude smaller with the deep learning methods compared to \sexp. Both network architectures achieve a comparable fraction of $\sim0.1\%$ outliers whereas the \sex fraction is of the order of $\sim10\%$, even when restricting the results to the cases where \sex detects both objects (see top panel of Figure~\ref{fig:one2one_nondet}). 

Lastly, as shown on the bottom panels of Figure~\ref{fig:one2one_nondet}, the performance of both \btof and \btomtof models on the galaxies that \sex did not manage to accurately deblend (25\%) gets affected compared with the well deblended cases (top panels) but remains unbiased with a low scatter and an outliers rate below $0.4\%$.

\begin{figure*}
\includegraphics[width=\textwidth]{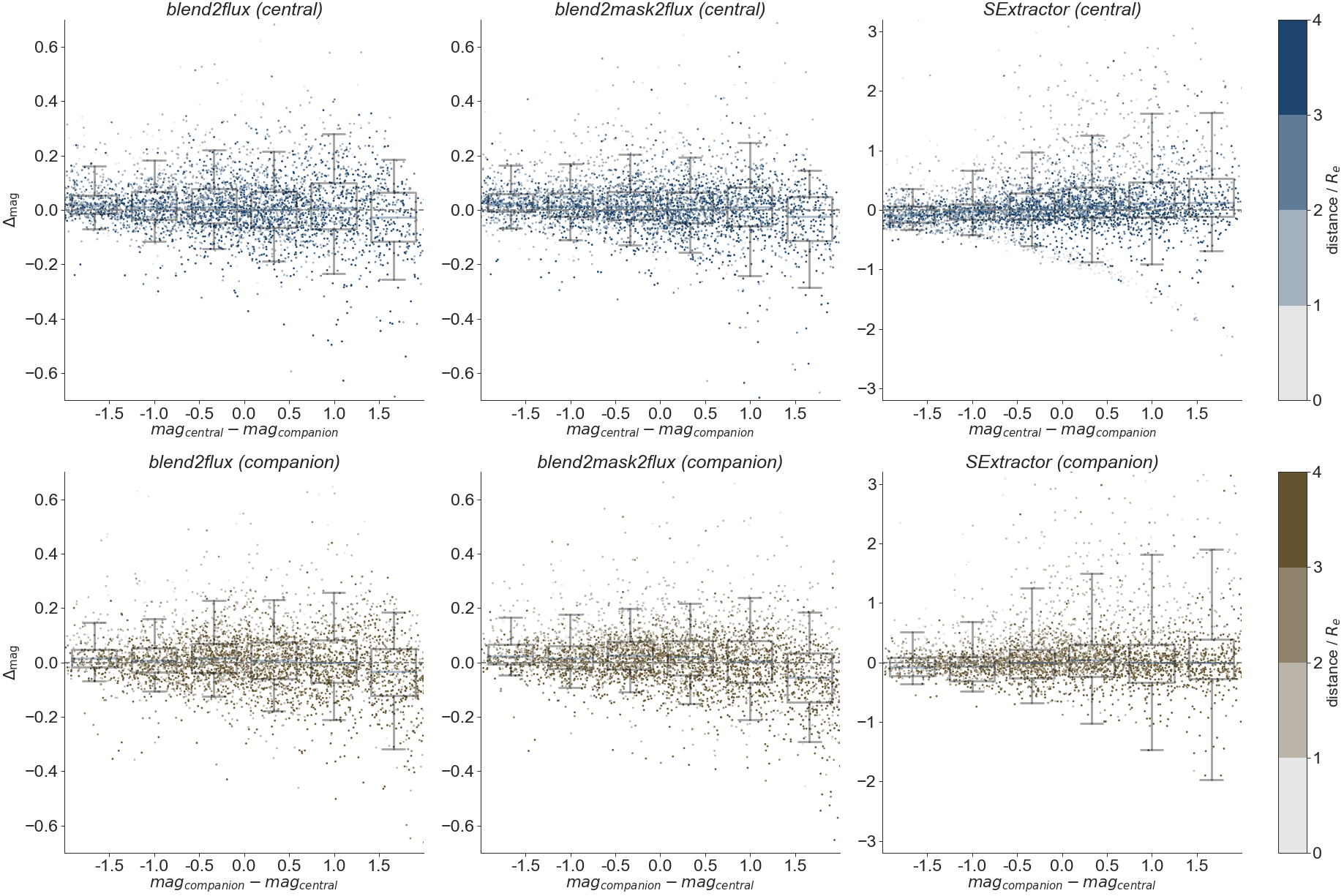}
\caption{Magnitude difference ($\Delta_{\rm mag}$) between the same galaxies when they are isolated (input) and blended (output) as a function of the magnitude difference in the blended system (${\rm mag_{central}-mag_{companion}}$). The top row shows the difference for the central galaxy. The bottom row corresponds to the companion galaxy. The columns indicate different codes. From left to right: \btof, \btomtof and \sexp. The boxplot marks the median and interquartile range (25\% - 75\%) for different bins in magnitude difference. The lines emanating from the box extend from $5^{\rm th}$ to $95^{\rm th}$ percentile of the data in each bin. The colour bar shows, for each blend, the distance between the objects normalised to the effective radius of the central galaxy.}
\label{fig:plts_dmag}
\end{figure*}

\subsection{Photometric accuracy vs. blend properties}

Aiming for an unbiased performance for a range of blend properties, we report results as a function of the magnitude difference between blended galaxies and the distance between the two objects.

In Figure~\ref{fig:plts_dmag} we show the magnitude difference between the isolated and blended galaxies 
(the bias in our magnitude estimate) as a function of the difference in magnitude between the two galaxies blended together. We observe that the two deep learning approaches present a very stable behaviour across the whole range of magnitude difference. As expected, the bias slightly increases when one of the galaxies in the pair is significantly brighter. However, it  remains below $\sim0.05$. Overall the bias remains always lower than the \sexp-based estimates. The deep learning results are also very stable in terms of scatter which is of the order of $\sim0.1$ magnitudes. Here the scatter for \sex based estimates is always significantly larger ($\sim 0.25$ magnitudes) than for the networks, and also shows a strong increase with magnitude difference between central galaxy and companion. For both networks this trend is only slightly visible.  

Figure~\ref{fig:plts_dmag} also encodes in the colour bar the normalised distance between the two galaxies.  Again, the deep learning results display little photometric dependence with distance, for both central galaxy and companion. The \sex results show a clearer dependence with distance, underestimating the fluxes up to 1 magnitude for close objects ($<R_e$). 

\begin{figure}
\includegraphics[width=\columnwidth]{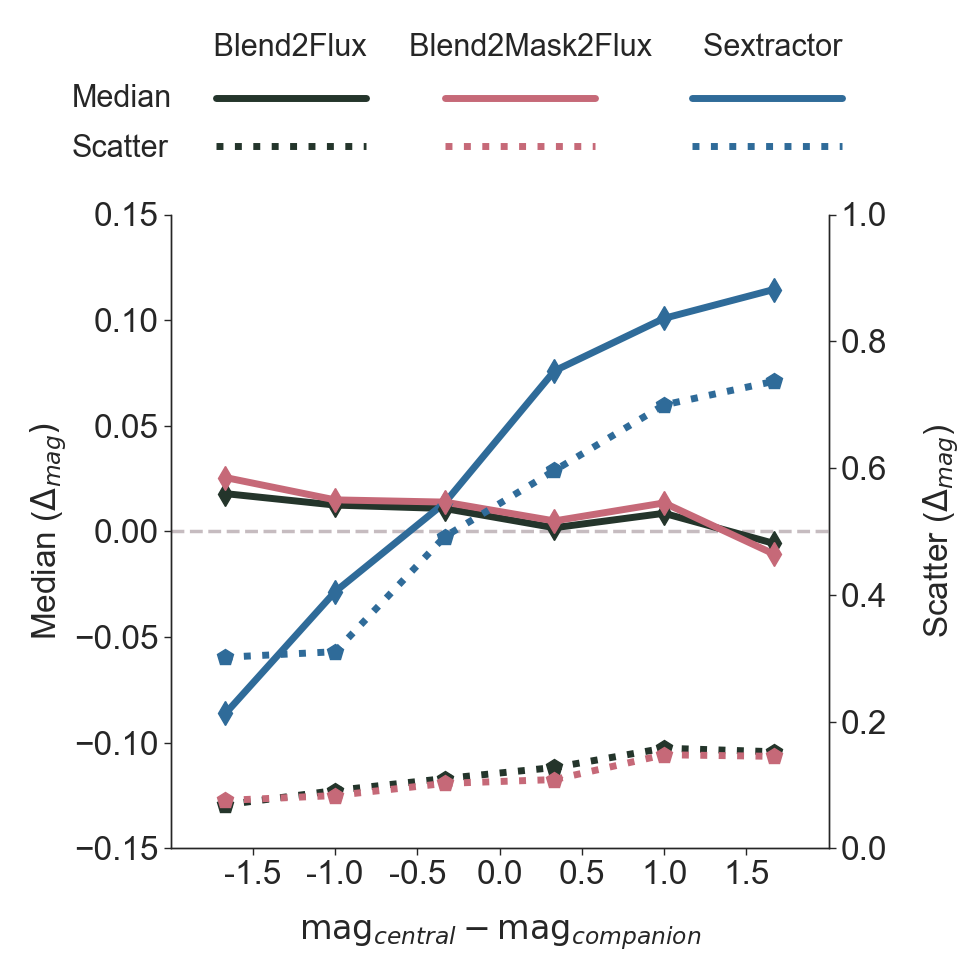}
\caption{Bias (solid lines) and scatter (dotted lines) in the recovered photometry in bins of magnitude difference for the central galaxy. The blue lines show the results for \sexp, the black lines correspond to the \btof configuration and the red lines to the \btomtof network. The bias value is reported on the left axis while the scatter value is reported on the right axis.
}
\label{fig:plts_cods_mdif}
\end{figure}

These trends are summarised in Figure~\ref{fig:plts_cods_mdif} which reports the bias and the scatter as a function of magnitude difference between the two galaxies. The \sex measurements are systematically more biased and more scattered than the machine learning-based estimates across the full range of parameters. We also notice that both architectures behave very similarly. 

\subsection{Photometric accuracy and morphology}

\begin{figure*}
\includegraphics[width=\textwidth]{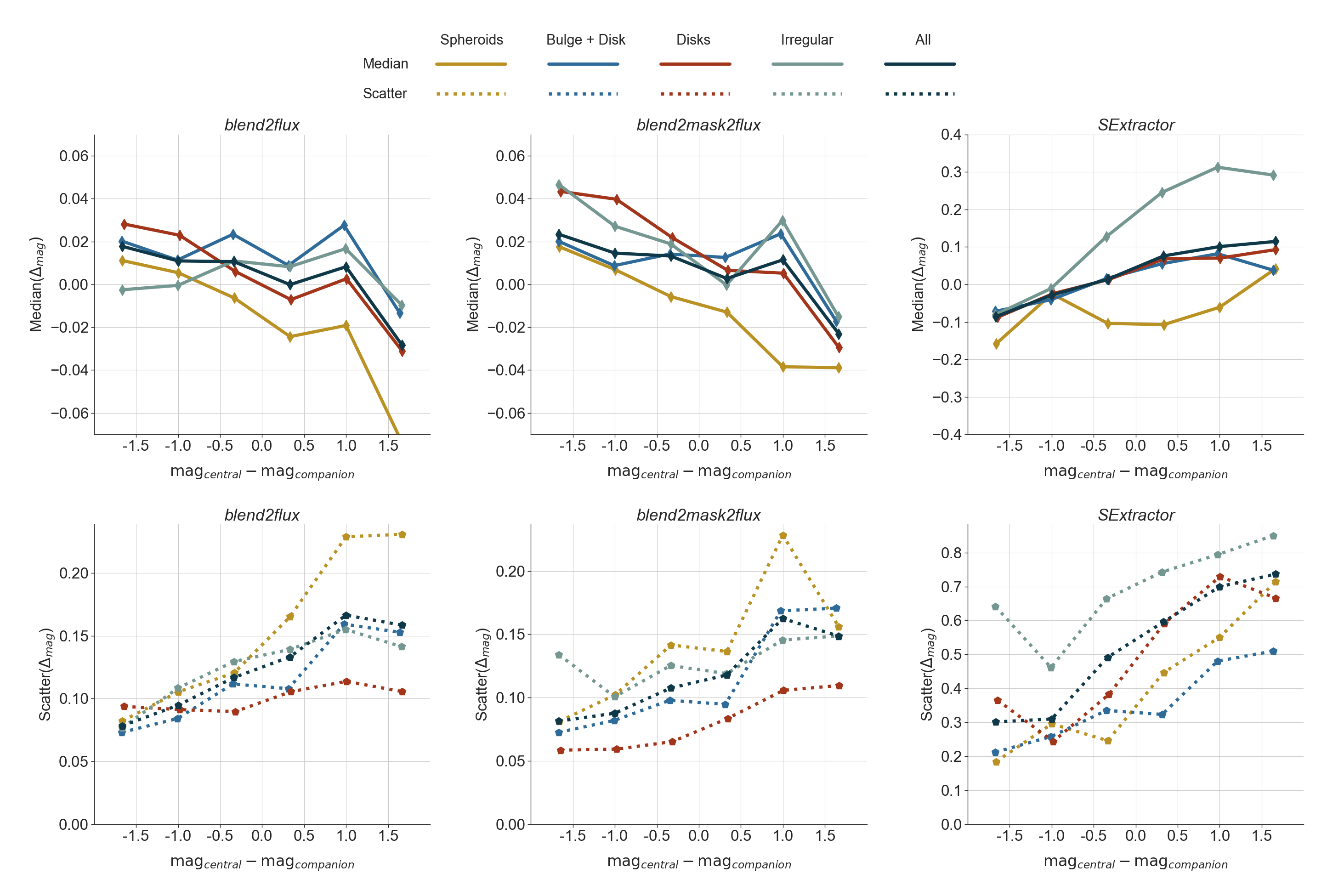}
\caption{Dependence of photometric bias (solid lines, top panels) and scatter (dotted lines, bottom panels) on the morphological type for the three codes considered in this work as a function of the magnitude difference. From left to right, the different panels show the results for \btof, \btomtof and \sex respectively. The different colours indicate the morphological type: spheroids (yellow), disk+spheroids (blue), disks (red) and irregulars (light green). The dark blue lines show the results for all galaxies.}
\label{fig:plts_morph}
\end{figure*}

The galaxies in our sample are classified into four morphological types (pure bulge, pure disk, two component bulge + disk, irregular) and are distributed as was shown in Table~\ref{tab:galprops}. One major property of the machine leaning methods presented here is that they do not assume any prior on the galaxy shape (as opposed to model fitting techniques). 
We explore in Figure~\ref{fig:plts_morph} the dependence of the photometric accuracy on the morphological type. We plot the median bias and scatter in bins of magnitude and distance now divided by morphological type. In general, the machine learning approaches show little dependence on performance with respect to morphology. As expected, irregular galaxies are harder to measure, and hence present a marginally larger scatter from both codes. Surprisingly, spheroidal galaxies tend to present larger errors when these galaxies are fainter than the other galaxy in the blended system ($\Delta_{\rm mag}>0$).
This behaviour seems to be present in both codes. The reason for this is unclear. One possible explanation is that the outskirts of the spheroids are too faint to be detected. Since these objects typically have large Sersic indices (i.e. steep luminosity profiles), the fraction of light in the outskirts is not negligible. \sex presents similar trends but overall more dramatic. In particular, the bias in the photometry of irregular galaxies is $\sim0.2$ larger than for the whole population. Also the luminosity of spheroids is systematically underestimated.  

As can be seen in Figure~\ref{fig:plts_morph} the photometric accuracy (magnitude scatter) overall is considerably lower for our two deep learning algorithms than for \sex results. 

\subsection{Segmentation maps}
\label{sub:segmaps}

\begin{figure*}
    \centering
    \includegraphics[width=0.9\textwidth]{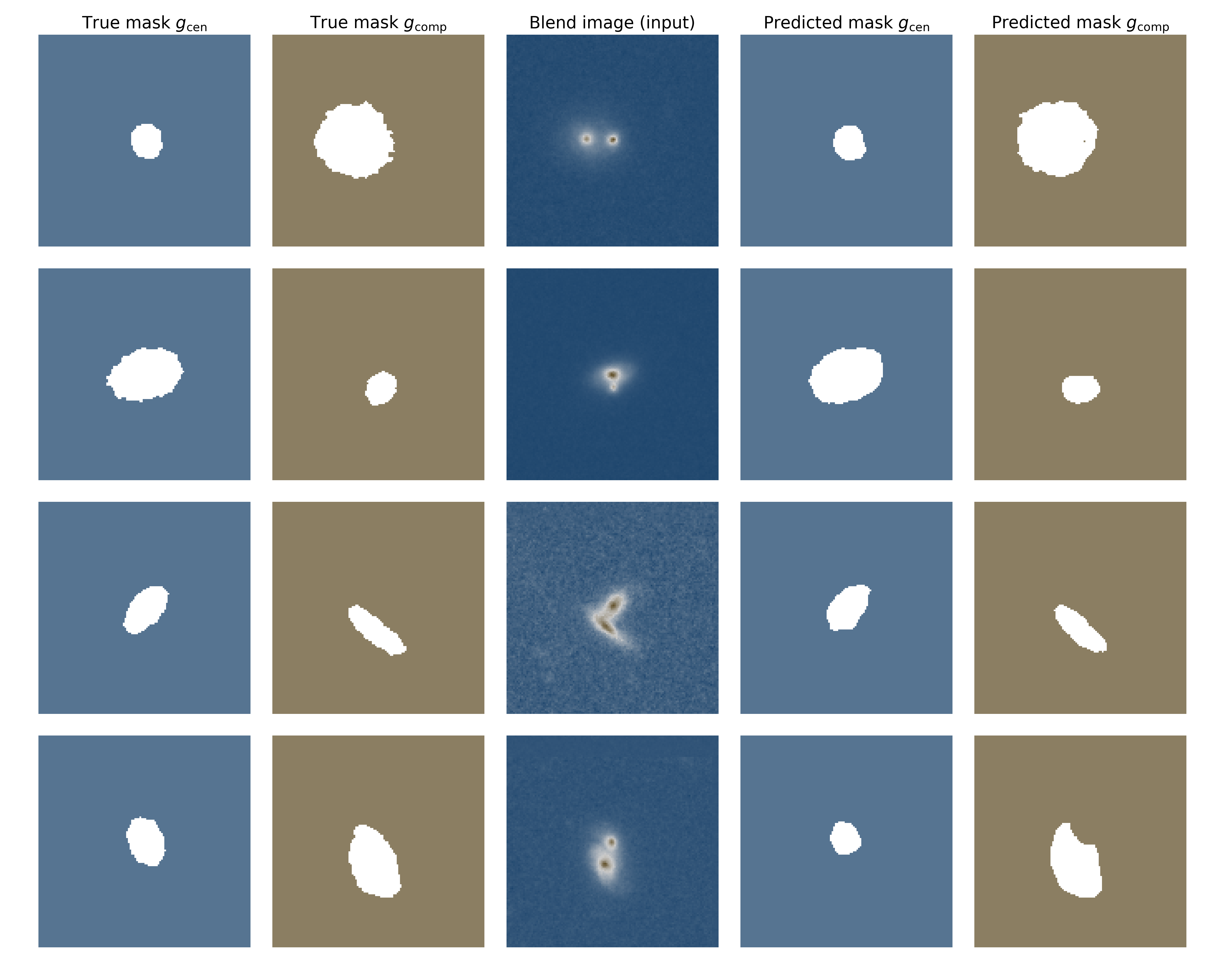}
    \caption{Selection of four simulated blends from the test data set and the recovery of the individual galaxy masks through the \btomtof network. At the centre is the stamp of blended galaxies that is input of the network. On the left are the segmentation masks obtained on the individual galaxy images with \sexp, and on the right the segmentation masks recovered by the network out of the blend image.}
    \label{fig:b2mmasks}
\end{figure*}

Throughout the paper, segmentation maps have been considered as a by-product of both \sex and the \btomtof network, possibly improving the photometry. In this subsection, we focus on the recovery of the segmentation maps of blended galaxies from the deep learning architecture, as well as the comparison between the results of the initial training of the \unet alone and the ones after the training of the full \btomtof network, which is characterised by the tuning of the segmentation maps for photometry.

\begin{figure*}
	\centering
	\includegraphics[width=0.9\textwidth]{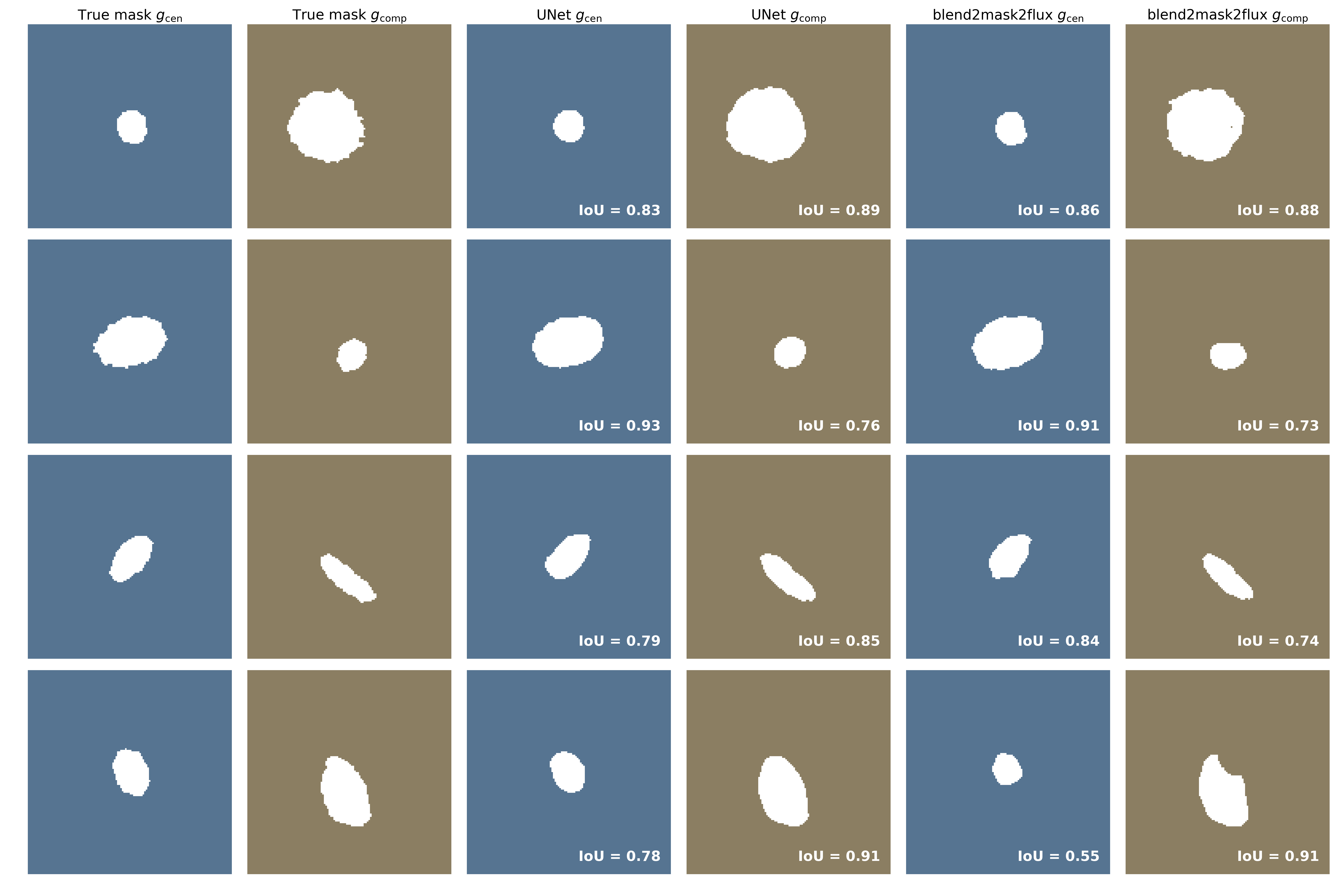}
    \caption{Same selection of four blend cases as in Figure~\ref{fig:b2mmasks} to compare this time only the results of the converged predicted segmentation masks of blended galaxies, yielded respectively by the \unet architecture (centre) and the \btomtof model (right), to the segmentation maps obtained from \sex on the individual galaxies. For each recovered galaxy mask, the segmentation score (IoU) as compared the \sex mask is indicated in the lower right corner.}
  \label{fig:segmaps}
\end{figure*}

In Section~\ref{sub:b2m}, we describe the \btomtof model as a hybrid network made of a \unet whose output is fed to a modified \btof network. The \unet is in charge of reproducing the two \sex segmentation maps of the original CANDELS galaxy cutouts from the blend image. In other words, its task is to take as input the full $128\times128$ blend image and produce two binary $128\times128$ images that correspond to the masks of the central and companion galaxy; this can be seen for a selection of four blends in Figure~\ref{fig:b2mmasks}. For better assessment of the accuracy of the method, we trained the network to output the segmentation maps in a specific order, central galaxy first, and then the companion. The cost function (loss) used to train the modified \unet is a \textit{binary-crossentropy}, which performs well for a pixel-by-pixel binary classification as needed for our segmentation maps. 

To score the results, the \textit{binary-crossentropy} loss is not very informative since every pixel rightfully classified as \textit{background} adds up to the accuracy, while we would like to assess the similarity to the target \sex segmentation map. For that purpose, we use a metric called Intersection over Union (IoU) also known as Jaccard index \citep{iou1901} of set A and B
\begin{equation}
    {\rm IoU}(A, B) = \dfrac{|A \cap B|}{|A \cup B|} = \dfrac{|A \cap B|}{|A| + |B| - |A \cap B|} .
\end{equation}
It is usually defined in computer vision for bounding boxes, but can be adapted to any shape. This metric has the advantage of decreasing very rapidly to zero in case of a mismatch between $A$ and $B$ in terms of location or morphology. Therefore a score superior to $0.5$ is considered a good score. 

After training, the \unet with the setup described in Section~\ref{sub:b2m} obtains an average IoU score of $${\rm IoU_{\unet}} = 0.82$$ on the test images, which is an indication of a very good recovery. 
However, once the \btomtof is trained end-to-end to recover the photometry, thus allowing 
the parameters of the \unet section to vary, the average IoU score on the test data set drops to $${\rm IoU_{\btomtof}} = 0.70$$.

The outcome and evolution from the pure segmentation objective to the photometry objective can be seen in Figure~\ref{fig:segmaps}, where the selection of blends is the same as the one on Figure~\ref{fig:b2mmasks}, but the segmentation results of the initial \unet are shown in the middle and can be directly compared with the one of the \btomtof model on the right. The IoU computed on each image with respect to the \sex segmentation on the left is indicated on each image.

\begin{figure*}
\begin{tabular}{cc}
\includegraphics[width=0.45\textwidth]{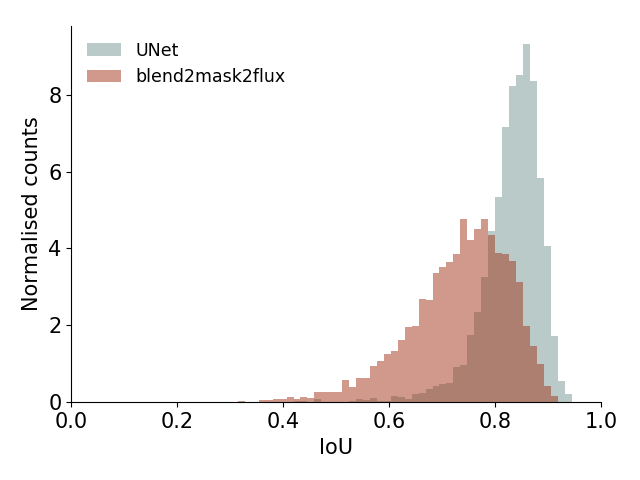} &
\includegraphics[width=0.45\textwidth]{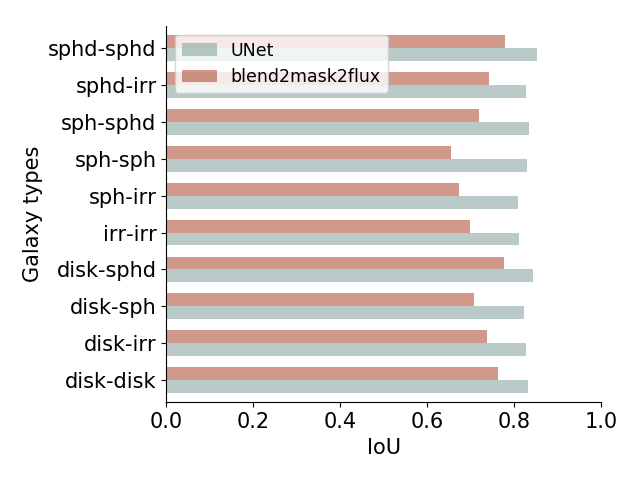}
\end{tabular}
\caption{Results of segmentation score (IoU) comparing the segmentation maps of the individual galaxies with the one obtained on the blended image with the \unet and the \btomtof model. On the left panel, we show the histogram of the distribution of the IoU values for all the blended galaxies in the test set. On the right panel we display the averaged IoU score on the test set as a function of the morphology of the galaxies in the blend. The \unet model being the starting point of the \btomtof model, we can consider the difference between the orange histograms and bars to the blue ones as an indicator of the impact on masks when optimising for the photometry.}
\label{fig:iou_results}
\end{figure*}

To further investigate the evolution of the segmentation when tuning the network for the photometry, we show the IoU statistics of the two models in Figure~\ref{fig:iou_results}. The left panel shows the histograms of the IoU score for both models. We see that the pure segmentation network has a very small dispersion, which broadens and becomes worse when it is optimised for the photometry. At this point, a possible interpretation would be that the target segmentation maps issued by \sex may not be fully adequate for flux measurements. To refine this claim, on the right panel, the IoU statistics are computed with respect to the morphology of the galaxies composing the blends. Each horizontal bar from the plot indicates for each model the average IoU score for the given pair of morphologies (no distinction is made which galaxy is central and which is companion galaxy). From this plot we see that, while the original \unet recovers the segmentation with the same accuracy throughout the entire spectrum of morphologies, the \btomtof network has more difficulties to recover the segmentation of certain types of morphologies, the irregular and the bulgy galaxies. While this is not very surprising for irregular galaxies, the fact that the segmentation of bulgy galaxies seems more difficult might help explaining the high variance that we see in the recovery of the photometry of these galaxies in Figure~\ref{fig:plts_morph}.

\section{Summary and conclusions}
\label{sec:conclusion}

We have presented a deep learning method to measure the photometry of blended systems in monochrome space-based images of the distant Universe. Firstly, we built a realistic training set out of observed high-redshift galaxies from the CANDELS survey which have been artificially blended. The data set covers a representative range of morphologies (bulges, disks, irregulars), magnitude differences ($-2<\Delta_{\rm mag}<2$) and distances ($0.5R_e<D<4R_e$) between the pairs. The data set of blended pairs is made public with this work to promote comparisons with other approaches.

We have tested two different neural network architectures. The first one measures the fluxes of the two galaxies directly from the images. The second approach, more complex, also estimates the segmentation maps. The networks are trained with a sample of $25,000$ galaxy pairs and tested on an independent sample of $5,000$ pairs. The results are compared to the standard \sex approach on the same test set. Our main results are:

\begin{itemize}
    \item Both deep learning approaches result in an unbiased photometric estimate with a typical uncertainty of $\sim0.1$ magnitudes. 
    This represents an improvement of at least a factor of 4 in flux error as compared with \sex even if the comparison is restricted to the cases where \sex detects exactly two objects as expected.
    \item The fraction of galaxies for which the photometric error exceeds 0.75 magnitudes is as low as $\sim1\%$ in the two machine learning approaches. This value reaches $\sim 12\%$ for \sexp. Our deep learning methods also reach an excellent photometric accuracy in these cases where \sex over- or under-deblends (i.e. finds more or less than two galaxies per image).
    \item The photometric accuracy obtained with the two deep learning approaches is very stable across all magnitude differences and distances explored in this work. Even for large magnitude differences between the two galaxies (factor $\sim 2$), the photometric uncertainty stays close to $\sim0.2$ magnitudes. This represents a major improvement as compared to \sex whose performance strongly depends on the properties of the blended system; at large magnitude differences its photometric uncertainty can reach 1 magnitude.
    \item The presented method does not assume any pre-defined model for the shape of galaxies. This is translated into a comparable photometric accuracy for all the morphological types explored in this work (disks, bulges and irregulars).
    \item Estimating and using the segmentation maps to estimate photometry results in a marginal gain in terms of photometric accuracy. The more complex network reaches slightly lower photometric errors and a smaller fraction of outliers at the expense of significantly larger training times. However, it has the advantage that the segmentation maps can be used to estimate uncertainties. We will explore this in future work.
    \item When a network is asked to optimise both for segmentation masks and photometry simultaneously, the recovered masks are usually tighter than the original ones derived by \sex on the isolated galaxies. This is especially true for irregular and bulgy galaxies.
\end{itemize}

This proof-of-concept work shows that machine learning can be used as a powerful tool on large imaging data sets, to measure the photometry. Despite the simplistic constraints we imposed on our dataset: two galaxies per stamp, one galaxy pinned at the image centre and no blends with completely overlapping galaxy centroids (also referred to as \textit{unrecognised blends}), our photometric measurement networks have demonstrated that on monochromatic images, they outperform traditional approaches with respect to photometric accuracy, precision, outliers fraction and stability towards different morphological types.

On top of that, we trained a network to also produce probabilistic maps of the presence of each individual galaxy. With a lower number of free parameters, the network using these maps systematically achieved better results on the photometry than the direct mapping between the blend image to the flux measurement. These probabilistic maps - that once thresholded are called segmentation maps - may well be used as a starting point by other software to guide the modelling of the blend galaxies.

Future work will focus on generalising the approach presented here to a more realistic situation, including multiple (>2) galaxies and more complex blend configurations.

\section*{Acknowledgements}

We acknowledge the support from CCIN2P3\footnote{\href{https://cc.in2p3.fr}{https://cc.in2p3.fr}} which provided the computing resources needed to train the deep learning models. Figures \ref{fig:seqstack} and \ref{fig:unet} made use of modified versions of the {\sc latex} package {\sc plotneuralnet} \citep{haris_iqbal_2018_2526396} and several code snippets openly available at \href{stack overflow}{https://stackoverflow.com}. We deeply thank everyone who contributes to open forums and learning platforms.

This work was partially developed during the $\rm 5^{th}$ COIN Residence Program\footnote{\href{https://cosmostatistics-initiative.org/residence-programs/coin-residence-program-5-chania-greece/}{https://cosmostatistics-initiative.org/residence-programs/coin-residence-program-5-chania-greece/}} (CRP\#5) held in Chania, Greece in September 2018. We thank Vassilis Charmandaris for encouraging the accomplishment of this event. COIN and EEOI are financially supported by CNRS as part of its MOMENTUM programme over the 2018 -- 2020 period. RSS acknowledges the support from NASA under the Astrophysics Theory Program Grant 14-ATP14-0007. 

\appendix




\bibliographystyle{mnras}
\bibliography{ref} 







\bsp
\label{lastpage}
\end{document}